# Enhancing Data Security in Medical Information System Using the Watermarking Techniques and Oracle SecureFile LOBs

Said Aminzou[1], Brahim ER-RAHA[2], Youness Idrissi Khamlichi[3], Mustapha Machkour[4], Karim Afdel[5]

[1, 2] Laboratory of Industrial and Computer engineering, ENSA Agadir, Morocco
[3] Department of Computer Engineering, ENSA of Khouribga
[4, 5] Laboratory of the Computing Systems and Vision, Faculty of Sciences Agadir, Morocco

[1] amnzou.said@gmail.com , [2] erraha@gmail.com , [3] ykhamlichi@gmail.com , [4] machkour@hotmail.com , [5] karim.afdel@gmail.com

**Abstract.** In this paper, we propose an efficient digital watermarking scheme to strengthen the security level already present in the database management system and to avoid illegal access to comprehensive content of database including patient's information. Doctors diagnose medical images by seeing Region of Interest (ROI). A ROI of a medical image is an area including important information and must be stored without any distortion. If a medical image is illegally obtained or if its content is changed, it may lead to wrong diagnosis. We substitute the part out of ROI of LSB bitplane of the image with the patient data and a binary feature map. This latter is obtained by extracting edges of the resized image to the quarter of its size using Laplacian operator. This image is directly integrated into the database. The edge map and invariant moments are used to check the integrity of the image.

## 1  Introduction

With the changes in the computer world and the health data security, securing medical information system is becoming a priority for citizens and for government.

Data security has three main properties: confidentiality, integrity and availability. Overall, the property of confidentiality prevents illegal access. The property of integrity guarantees detection of any modification of data, whether accidental or malicious. Finally, the property of availability protects the system against the attacks of denial of service.

Digital watermarking is a technology for embedding digital information in digital content (audio, images, video…) [1][2][3]. It has introduced as a tool to improve the security, and different watermarking schemes have been proposed to address the problems of medical privacy and security [4][5][12].

In this article we present a secured system for medical database. The security mentioned in this paper is primarily concerned with the integrity and confidentiality of data. The section 2 includes a description of the software that was used for the implementation of our system, special attention will be reserved for the ORACLE Database Management System and security procedures offered by the latter to ensure confidentiality. The section 3 describes the functionalities of this system in particular the one of watermarking used like an additional mean to enhance the integrity and confidentiality in database system.

## 2  Environment of achieving system

To develop our system we use Java language and Oracle DBMS (Database management system). Java was used to design the user interface and implement the functionalities of the system. The Oracle database was used for the storage and retrieval of data.

### 2.1  Java

We used Java to develop the user interface on the one hand, and implement the functionalities of the system on the other.

We chose this language for security, portability and particularly his API (Advanced Programming Interface) with classes of handling of image.

**2.2  The Oracle DBMS with SecureFiles**

Oracle is a database system that can manage complex objects like image. In previous versions of Oracle, The image is either integrated into the database (BLOB) or outside (BFILE) storing a descriptor of the file containing this image (Figure 1). [6]

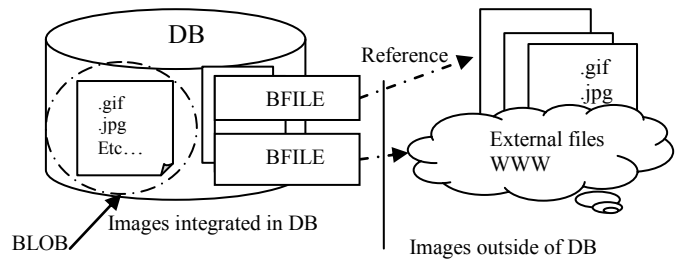

**Fig. 1.**  Image BLOB and BFILE

In this work, we use the new Oracle Database 11g technology, Oracle SecureFiles.

Oracle SecureFiles is a novel architecture that provides the most scalable execution of filesystem-like unstructured content and advanced filesystems while preserving the rich features and benefits of the Oracle database server [7].

SecureFile LOBs allow the integration of images in the database and hence benefit from:

- Super fast and powerful file storage.
- Sharing, confidentiality, consistency, etc.

**2.3  Interaction Java / Oracle**

The integration of SQL code in the Java code allows the manipulation of textual data from the database (information about the patient) using the user interface. This manipulation is provided by the JDBC interface. In addition, with the ORDImage class (Java class) we can perform basic operations on images SecureFile LOBs (insertion, extraction and modification).

## 3   Functionalities of the system

Our system includes the basic functionalities of any information management system to insert, view and update data.

In addition, it should allow the practitioner to achieve the following:

- Construction of the watermarked image,
- Storage of the watermarked image,
- Verify the integrity of the medical image,
- Retrieve patient information from the watermarked image,
- Archiving original image.

### 3.1   Construction of the watermarked image

To ensure the integrity of images and confidentiality, we used the technique of watermarking where information and images of the patient are a single entity. The construction of the watermarked image is one of the main functions of our system. It includes the following tasks:

- retrieve the original image from database,
- separation of the LSB plane[1] of the image (Figure 2),
- construction of signature with the moments [8] of the edge map[2] (Figure 3),
- encryption of patient data and signature,
- integration of encrypted data in the LSB,
- construction of the watermarked image.

The process of achieving this function is shown on a mammogram in Figure 2. The information is entered in first step, stored in the database without any treatment and in particular on the image. In the second step, to get watermarked image, we proceed to the integration of the ROI part of LSB and patient information and our signature using edge map of the image [9][10][11] and invariant moments [12]. The resulting image is then stored in database. For reasons of security and performance of the database, the original image will be saved on an external device: magnetic tape, DVD, etc.

---

[1] Least Significant Bits plane: the back plane of the image.
[2] map of points representing contour's image.

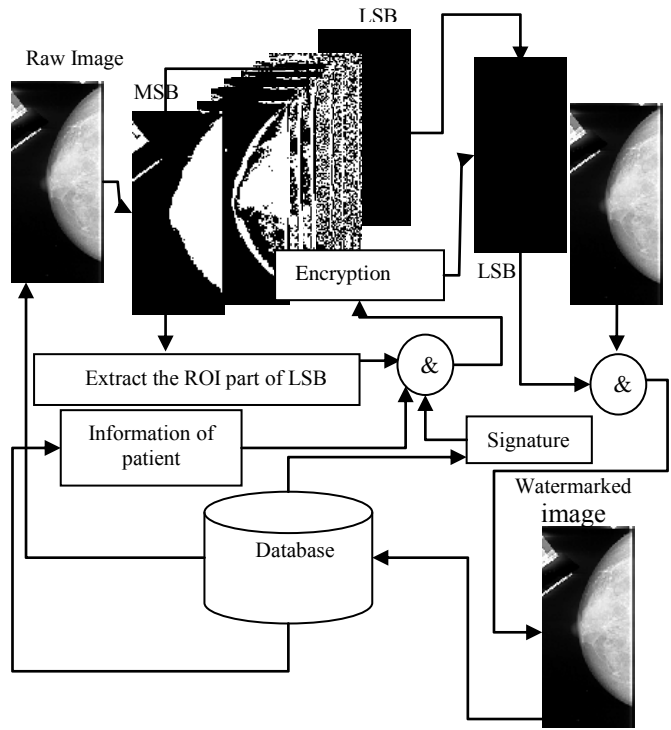

**Fig. 2.** Process of construction and storage of the watermarked image.

### 3.1.1 Construction of the signature

The signature is composed of both the edge map and the average value of functions of moments. On the one hand the moments of second and third order are known to be invariants of the image. In other words, the values of these moments remain unchanged during a change in scale [13], rotation and / or orthogonal transformations [10][14][15]. On the other hand, the edge map uniquely characterizes the image. Any intentional or accidental modification of the image of the patient affects this map.

### 3.1.2  Creating the edge map

The edge map is obtained by applying the operator Laplacian of Gaussian (LOG) [16] on the original image without LSBs bitplane. To increase the ability to integrate and reduce the space occupied by this map in terms of LSB bits of the image, we have reduced by a factor of 4 the size of the original image before the creation of this map. The logic of using the edge map is that any change on the image will affect the edge and result in the alteration of this map. This map is also used to locate the modified area. Figure 3 shows the steps for creating the edge map of a mammography image.

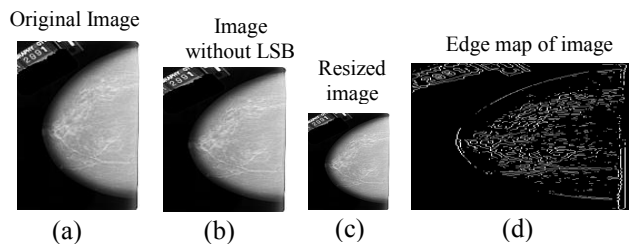

(a)      (b)      (c)      (d)

**Fig. 3.**   Creating the edge map of a mammography image

### 3.1.3  Integration of the signature

The first step is to calculate the average value of the moments of second and third order on the image of mammography without LSBs bitplane.

The second step begins with the extraction of the edge map (Figure 4 (a)), which will be subdivided into blocks of 6x6 pixels. These blocks are divided in terms of LSB bits to make the difficult reconstruction of the map to illegal access. The technique of arrangement consists to replace the right part of the block by the left part of the neighbouring block and this along a circular well definite way (Figure 4 (b)).

The LSBs bitplane of the image is then replaced by the ROI part of LSB and encrypted data consists of information of the patient, the average value of the moments of the second and third order, and blocks of edge.

We used the symmetric AES (Advanced Encryption Standard) algorithm [17] for data encryption.

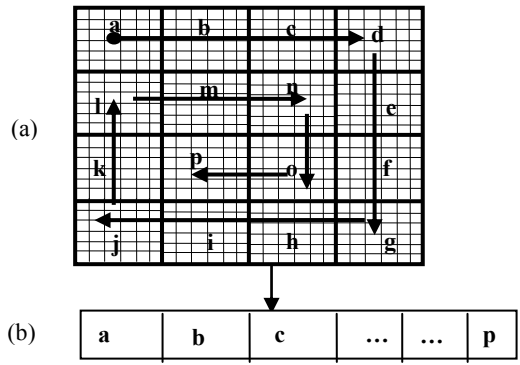

(a)

(b)

**Fig. 4.** Rearranging the edge map in blocks of size 6x6 pixels.

(a) The edge map is divided into blocks alphabetically ordered along a circular path;
(b) The blocks are arranged continuously.

### 3.2  Locating of tamper region

The verification of the image integrity is a crucial function in our system. To check the integrity of an image already stored in the database, we begin by extracting patient's data and the average of moments (above cited) from the LSBs bitplane of this image. Then we recalculate moment invariants of the image without the LSBs bitplane. If the calculated moments is identical to that extracted from the LSBs bitplane, then the retrieved image is intact. Otherwise, the image has been altered and its integrity is not preserved. This process is illustrated in Figure 5.

   In the case where the image is altered, we can identify the altered region. We rearrange the edge map integrated in the LSBs bitplane by reversing the steps described above. We then compare the results to edge map of the retrieved image without the LSBs bitplane. A simple comparison of two maps allows localizing the affected region (see Figure 5).

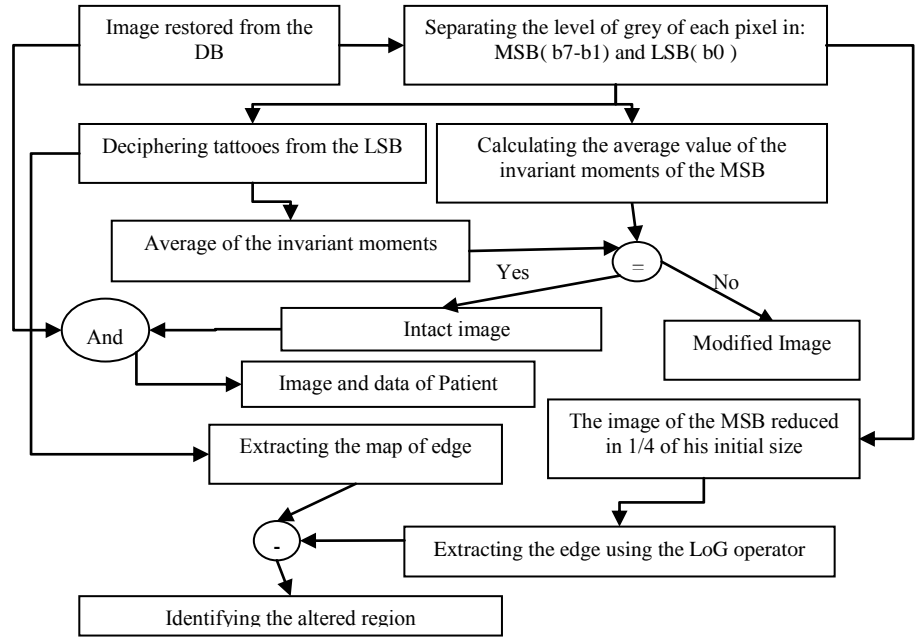

**Fig. 5.** The process of verifying the integrity of the image.

This verification process is also shown in Figure 6 with experimental values on a mammography image.

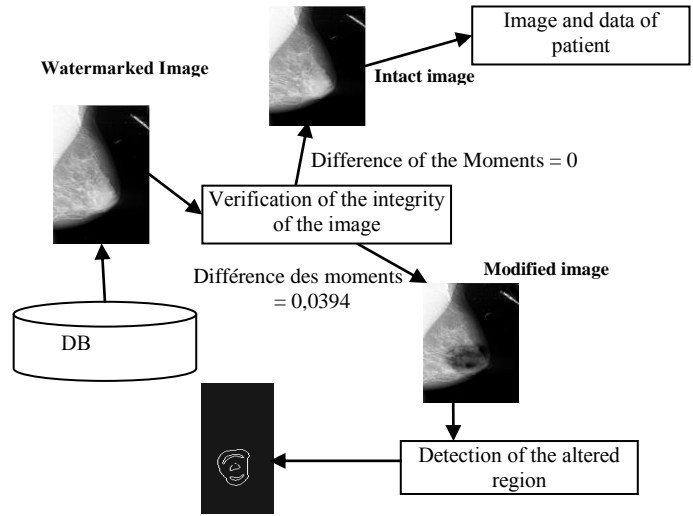

**Fig. 6.** The process of verifying the integrity of a mammography image.

## 4  Conclusion

In this article we have implemented a security architecture using watermarking and encryption techniques in addition to the security provided by database management system. As proprieties of the security, we have treated the confidentiality (Figures 2, 3 and 4) and integrity (Figures 5 and 6) of both data and image.